# A change-point detection method for detecting and locating the abrupt changes in distributions of damage-sensitive features of SHM data, with application to structural condition assessment

Xinyi Lei[1,2,3], Zhichng Chen[1,2,3], Hui Li[1,2,3], Shiyin Wei[1,2,3]

[1]Key Lab of Intelligent Disaster Mitigation of the Ministry of Industry and Information Technology, Harbin Institute of Technology, Harbin, 150090, China

[2]Key Lab of Structures Dynamic Behavior and Control of the Ministry of Education, Harbin Institute of Technology, Harbin, 150090, China

[3]School of Civil Engineering, Harbin Institute of Technology, Harbin, 150090, China

**Abstract**

Diagnosing the changes of structural behaviors using monitoring data is an important objective of structural health monitoring (SHM). The changes in structural behaviors are usually manifested as the feature changes in monitored structural responses; thus, developing effective methods for automatically detecting such changes is of considerable significance. Existing methods for change detection in SHM are mainly used for scalar or vector data, thus incapable of detecting the changes of the features represented by complex data, e.g., the probability density functions (PDFs). Detecting the abrupt changes occurred in the distributions (represented by PDFs) associated with the damage-sensitive features extracted from SHM data are usually of crucial interest for structural condition assessment; however, the SHM community still lacks effective diagnostic tools for detecting such changes. In this study, a change-point detection method is developed in the functional data-analytic framework for PDF-valued sequence, and it is leveraged to diagnose the distributional information break encountered in structural condition assessment. A major challenge in PDF-valued data modeling or analysis is that the PDFs are special functional data subjecting to nonlinear constraints. To tackle this issue, the PDFs are embedded into the Bayes space, and the associated change-point model is constructed by using the linear structure of the Bayes space; then, a hypothesis testing procedure is presented for distributional change-point detection based on the isomorphic mapping between the Bayes space and a functional linear space. Comprehensive simulation studies are conducted to validate the effectiveness of the proposed method as well as demonstrate its superiority over the competing method. Finally, an application to real SHM data illustrates its practical utility in structural condition assessment.

## 1. Introduction

Structural health monitoring (SHM) is an important research field in civil engineering; it involves installing sensors on important buildings and infrastructure (e.g., bridges, tunnels, dams) to obtain a real-time monitoring for structural responses, operational environments, internal and external loads, then performing condition assessment or health diagnosis for the structure based on analyzing and mining the monitoring data [1-4]. Detecting changes in the damage-sensitive features (DSFs) extracted from the monitoring data of interest is a very important content in SHM data mining [1-14], as it usually contributes to providing useful even critical information for further structural condition assessment or damage detection. So far, various change detection methods have been developed or adopted for different SHM applications; representative methods or techniques include change-point detection method [5-8], baseline reference method [9,10], parameter tracking technique [11], correlation tracking method [12], subspace-based fault detection method [13], and cluster analysis [14]. Moreover, in another closely related filed, namely the machinery health monitoring (MHM) in mechanical engineering, a series of change detection methods have also been developed for MHM signal processing and damage detection [15-21]. To date, the change detection methods employed for structural or machinery monitoring data mining are generally applicable to scalar or vector data. In other words, the detection methods can only handle the features that can be converted to be represented as scalar- or vector-valued data (usually in the form of real- or vector-valued time series) living in the Euclidean space, thus limiting their applications in detecting the changes in the features represented by complex data objects without Euclidean data structure.

The SHM data contain huge amount of information, rich information is carried by complex data (e.g., high dimensional data, functional data). By analyzing the complex data extracted from the monitoring data, one can uncover new information that cannot be clarified by conventional scalar or vector data analysis. One representative type of complex and non-Euclidean data that can be extracted from the monitoring data is the probability density function (PDF). For instance, if we split the monitoring data into a series of time-segments, the PDFs estimated from the segment data can form a new dataset, namely $\{f_i: i = 1,2,\cdots,n\}$ with $f_i$ being a PDF. Generally, the extracted PDFs are continuous functions, which naturally belong to functional data. A directly related branch of modern statistics specifically for analyzing such data object consisting of random functions is the functional data analysis (FDA) [22, 23]. For detailed introductions to FDA, the readers are referred to the monographs of Ramsay and Silverman [22], Ferraty and Vieu [23]. The objective of this study is to develop an FDA-based methodology for detecting and locating the potential changes occurred in the PDF-valued sequence, namely $\mathcal{F} = \{f_i: i = 1,2,\cdots,n\}$, extracted from the monitoring data. For convenience, such a functional sequence consisting of random PDFs is also called distributional sequence throughout this study. The motivations are mainly two-fold: (1) on the one hand, the distributional change detection method can provide an automatic detection tool for diagnosing the distributional information breaks involved in structural condition assessment. The distributions associated with the feature variables extracted from structural responses play an important role in structural health diagnosis [2]. Researchers have found that performing condition assessment based on distributional information (carried by PDFs) is more effective than conventionally used vibration features (e.g., natural frequencies and mode shapes) in some situations [2, 24, 25]. So far, distribution-change-based structural condition assessment methods have been developed and successfully applied in SHM for local structural state diagnosis, namely stay cables and steel box girders of long-span bridges, see Li et al. [24] and Wei et al. [25] for more details. The abrupt change of the distributional features is the critical information for such structural condition assessment methods; however, the SHM community still lacks effective tools for automatically detecting and locating such distributional changes. (2) On the other hand, distributional change detection can provide important guidance for model switching for some related statistical methods dealing with distributional data employed for SHM applications. So far, several different distributional data-analytic methods (e.g., distribution-to-distribution regression [26, 27], functional principal component analysis (FPCA) for distributional data [27, 30]) have been successfully applied in SHM for different tasks, such as data reconstruction [26-30] and complex dependence analysis [31];

however, if the investigated distributional data changed, the corresponding statistical models should be adjusted or re-trained accordingly, otherwise it would lead to unreliable analyses or predictions.

From the statistical perspective, the distributional change detection method to be developed in this study falls into the category of change-point detection. Change-point detection originates from quality control, now it has become an important research topic in statistics [32-34]. Given a sequence of observations, namely $\{X_i: i = 1,2,\cdots,n\}$ generated by an unknown random process, the change-point detection problem concerns of detecting and locating the changes in the random process that governs the observations [33]. In some statistical literature, change-point detection is also called structural break detection (or analysis) [32]. To date, change-point detection has been widely used in various fields, such as engineering [5-8, 15-20, 38-41], finance [34], medicine [34], climatology [35], psychology [36], hydrology [37], and others. In the engineering discipline, in addition to the earlier-mentioned applications involved in structural or machinery health monitoring [5-8, 15-20], change-point detection has long been used in signal processing encountered in fault detection and adaptive process control [38-41].

In the field of statistics, change-point detection methods for scalar data have been developed for a long history[32, 33]; however, for functional data, related literature has only emerged in the last decade. Existing methods pertinent to functional change-point detection are mainly developed for ordinary functional data, the investigated data objects are generally assumed to lie in the $L^2[a,b]$ space (a type of linear space formed by all square integrable functions), representative studies include Berkes et al. [42], Aue et al. [43], Hörmann and Kokoszka [44], and Aston and Kirch [45]. Conventional functional change-point detection methods mainly rely on the FPCA technique to convert the functional version of the change-point problem to the scalar version [42-45]. As pointed out by Aue et al. [46], if the functional change is orthogonal to the FPC-basis where the data are projected onto, the detection results obtained by FPCA-based approaches might be unreliable. To overcome this weakness, Aue et al. [46] recently proposed a "fully functional" approach, free from the dimensional reduction processing, for functional structural break analysis, which effectively avoids the issue suffered by the FPCA-based approaches. However, PDFs are special functional data, not only subjecting to inherent constrains (i.e., $f(x) \geq 0, \forall x \in \mathcal{D}$ and $\int_{\mathcal{D}} f(\tau)d\tau = 1$ for arbitrary PDF $f(x)$ defined on the domain $\mathcal{D}$) but also not residing in a linear space under ordinary linear operations [47-49]. In addition to PDFs, distributions can also be characterized by cumulative distribution functions (CDFs) or quantile functions (QFs); however, such functions also have nonlinear constrains [47]. Such an inherently nonlinear nature makes statistical analysis for distributional data more challenging as they usually violate the existing linear methods developed for ordinary functional data [47,49].

So far, only a few research contributions on change-point detection for distributional data have been reported and only begun at the last few years. Padilla et al. [50] developed a nonparametric approach for distributional change detection, the distributions are assumed to be identical within two segments of the data sequence separated by the change-point. The detection procedure was developed based on windowed Kolmogorov-Smirnov statistics, and the robustness of the detection method with respect to the window size was also investigated. Most recently, Horváth et al. [51] proposed another novel change-point detection method for distributional sequence based on Wasserstein distance, where the convergence and asymptotic laws were comprehensively investigated. The distributions before (or after) the change-point are also assumed to be equal (i.e., $F_t = F$ when $t \leq k^*$, and $F_t = G$ ($\neq F$) when $t > k^*$, where $F_t$ stands for the distribution function and $k^*$ stands for the unknown change-point). Another excellent contribution is the Fréchet change-point detection method proposed by Dubey and Müller [52], such an approach is developed for data objects residing in a general metric space, and it is applicable to random PDFs and networks. However, for engineering applications, existing distributional change-point detection methods still have their imperfections. For instance, the identical distribution assumption made in Padilla et al. [50] and Horváth et al. [51] is too strong to be realistic for the distributional sequence extracted from SHM data. Additionally, existing methods generally lack effective measures to cope with the adverse impacts caused by outliers presented in the dataset. In fact, sensitive to outliers is a common issue that suffered by most of the existing change-point detection methods. It is worth noting that the SHM systems operate in complex environments, data contamination is very common and inevitable; thus, being able to provide reliable results under data contamination is a basic performance requirement for the detection methods employed for SHM applications.

This study proposes an alternative change-point detection method for distributional data employed for distributional information break detection in SHM applications. The distributions before and after the change-point are assumed to be respectively generated from two different functional random processes. In contrast to the identical distribution assumption made in Padilla et al. [50] and Horváth et al. [51], the distributions before and after the change-point in our method are two groups of distributions, and the elements are not necessary to be identical within the same group. The change-point model for the distributional sequence is built on the linear structure of the Bayes space after imbedding the PDFs into the Bayes space. The associated hypothesis testing procedure for change-point detection is developed by leveraging the isometric isomorphism between the Bayes space and $L^2$ space. To obtain more reliable detection results under data contamination, a functional data cleaning-based strategy is also presented to handle the outlying PDFs. To demonstrate the practical usefulness of the proposed method in structural condition assessment of bridge components, it is applied to diagnoses the distributional information break (a critical information for distribution-change-based structural condition assessment method [24]) of the cable-tension monitoring data from a real long-span bridge.

## 2. Distributional change-point detection method

This section presents the proposed change-point detection method for distributional sequence. Firstly, the change-point model of the ordinary functional sequence is briefly recalled in Subsection 2.1, along with discussions on its limitations in distributional data modeling; then, in Subsection 2.2, an alternative change-point model constructed in the Bayes space will be proposed specifically for distributional data, based on which the proposed distributional change-point detection method will be provided in Subsection 2.3. Finally, the last subsection gives the measures for ensuring the reliability of detection under data contamination.

*2.1 Preliminaries*

This subsection briefly introduces the mean break (i.e., abrupt change in the mean) model for ordinary functional data described in Aue et al. [46]. Consider a functional sequence consisting of $n$ random functions taking values in the $L^2[a,b]$ space, namely $\{X_1(t): i = 1,\cdots,n\}$ with $X_i \in L^2[a,b]$ and $t \in [a,b]$, the mean break of this functional sequence can be modeled as

$$X_i(t) = \mu(t) + \chi_{[k^*+1,\,n]}(i)\delta(t) + \varepsilon_i(t), \quad t \in [a,b] \tag{1}$$

where $\mu(t)$ is the baseline mean function, $\delta(t)$ is the increment of the mean function, $k^*$ is the change-point location, $\varepsilon_i(t)$ is the zero-mean error term, and $\chi_{[k^*+1,\,n]}(i)$ is the indicator function defined as

$$\chi_{[k^*+1,\,n]}(i) = \begin{cases} 1, & \text{if } i \in [k^*+1,\,n] \\ 0, & \text{if } i \notin [k^*+1,\,n] \end{cases} \tag{2}$$

However, if this model is employed to characterize the mean break occurred in a PDF-valued sequence denoted as $\{f_i(x): i = 1, \cdots, n\}$ with $f_i$ being a random PDF, i.e.,

$$f_i(x) = \mu_f(x) + \chi_{[k^*+1,\,n]}(i)\delta_f(x) + \varepsilon_{f_i}(x), \quad t \in [a, b] \tag{3}$$

it would fail to ensure that $\mu_f$, $\delta_f$ and $f_i$ can simultaneously satisfy the constrains of PDFs (because the PDF space is not closed under ordinary linear operations, namely the pointwise addition and scalar multiplication defined for ordinary functional data). If the mean functions $\mu_f(x)$ and $\mu_f(x) + \delta_f(x)$ before and after the change-point are no longer PDFs, the associated distributional mean break model would become uninterpretable.

*2.2 Bayes space-based mean break model of distributional sequence*

The main challenge for PDF-valued data modeling is that the PDF space lacks the linear structure in usual sense. However, by introducing the following additive and scalar multiplication operations to PDFs [53-55]:

addition: $(f \oplus g)(x) = \dfrac{f(x)g(x)}{\int_I f(\tau)g(\tau)d\tau}$, $x \in I$ (4a)

scalar multiplication: $(c \odot f)(x) = \dfrac{f(x)^c}{\int_I f(\tau)^c d\tau}$, $x \in I$, $c \in \mathbb{R}$ (4b)

where $f(x)$ and $g(x)$ are both PDFs defined on the compact interval $I = [a, b]$, the PDFs can be regarded as the elements of the Bayes space denoted as $\mathfrak{B}^2(I) = \mathfrak{B}^2([a,b])$ (see the Appendix for the theoretical background and notions). The Bayes space is a linear space [54,55].

Next, by using the linear structure (i.e., the linear operations defined in Eq. (4)) of the Bayes space, the functional mean break model given in Eq. (1) will be extended to accommodate PDF-valued data. Considering a functional series composed of $n$ PDFs residing in the Bayes space $\mathfrak{B}^2(I) = \mathfrak{B}^2([a,b])$, i.e.,

$$\mathcal{F}(I) = \{f_1(x), f_2(x), \cdots, f_n(x)\}, \quad f_i \in \mathfrak{B}^2(I), \quad x \in I = [a, b] \tag{5}$$

this sequence can be treated as a realization of an unknown distributional stochastic process consisting of $n$ PDF-valued random variables. Without loss of generality, the common support $I = [a, b]$ of the PDFs is assumed to be $I = [0,1]$. For the PDFs supported on a general domain, namely $I = [a, b]$, they can be converted to be supported on $[0,1]$ through performing a scalar transformation, see, e.g., Dasgupta et al. [56] and Chen et al. [26-27] for more detailed descriptions. For notation simplification, $\mathcal{F}(I)$ will be shorted as $\mathcal{F}$ with $I$ dropped throughout the rest of this study. By this setting, the sample mean function of the $n$ PDFs can be calculated as

$$\hat{\mu}_f(x) = \left(\frac{1}{n} \odot \left(\bigoplus_{i=1}^{n} f_i\right)\right)(x), \ x \in [0,1] \tag{6}$$

It can be easily verified that $\hat{\mu}_f$ is an element of the Bayes space $\mathfrak{B}^2([0,1])$, because the latter is closed under the linear operations defined in Eqs. (4a) and (4b).

After embedding the PDFs into the Bayes space, the mean break model for ordinary functional data can be extended to model the distributional mean break as follows:

$$f_i(x) = \left(\mu_f \oplus \left(\chi_{[k^*+1,\ n]}(i) \odot \delta_f\right) \oplus \varepsilon_{f_i}\right)(x), \ x \in [0,1] \tag{7}$$

where $\mu_f$, $\delta_f$, $\varepsilon_{f_i}$ and $k^*$ have the analogous meanings as their counterparts given in Eq. (1). Before the change-point (i.e., $1 \leq i \leq k^*$), the mean function of the PDF-valued random variables in the distributional stochastic process is $\mu_f$; after the change-point (i.e., $k^* + 1 \leq i \leq n$), the mean function of the PDFs becomes $\mu_f \oplus \delta_f$. Recall that the Bayes space $\mathfrak{B}^2([0,1])$ is closed under the operations $\oplus$ and $\odot$, one can verify that $\mu_f$, $\delta_f$ and $\mu_f \oplus \delta_f$ can simultaneously satisfy the constrains of PDFs. In this sense, the new model constructed in the Bayes space is more interpretable for distributional mean break modeling.

*2.3 Change-point detection method*

This subsection presents the change-point detection method for the distributional sequence based on the Bayes space mean break model described above. The change-point detection problem concerns testing the existence of changes in the mean of the distributional sequence through hypothesis testing; if the change occurs, then the location of the change-point is further estimated. With the mean break model given in Eq. (7), the problem of change-point detection is equivalent to test whether the mean-increment $\delta_f$ is zero or not at a pre-specified significance level. In this sense, the null and alternative hypothesis for change-point detection can be written as

$$H_0: \delta_f = 0, \ H_A: \delta_f \neq 0 \tag{8}$$

Inspired by the functional cumulative sum statistic (FCSS) constructed in Aue et al. [46] for ordinary functional data defined in the $L^2([0,1])$, the following modified FCSS is constructed in the Bayes space as the basic statistic in subsequent change-point localization and significance test:

$$F_{n,k}^0 = \frac{1}{\sqrt{n}} \odot \left(\left(\bigoplus_{i=1}^{k} f_i\right) \oplus \left(-\frac{k}{n}\right) \odot \left(\bigoplus_{i=1}^{n} f_i\right)\right) \quad (9)$$

It is worth noting that $F_{n,k}^0$ also resides in the Bayes space $\mathcal{B}^2([0,1])$, and its $\mathcal{B}$-norm $\|F_{n,k}^0\|_\mathcal{B}$ (induced by the inner product defined in Eq. (24) of the Appendix, i.e., $\|f\|_\mathcal{B} = \sqrt{\langle f, f \rangle_\mathcal{B}}, \forall f \in \mathcal{B}^2([0,1])$) is a function of the index $k$. Similar to the property of the FCSS in the $L^2([0,1])$ sense [46], $\|F_{n,k}^0\|_\mathcal{B}$ also tends to reach its maximum value at $k = k^*$; thus, the change-point $k^*$ of the distributional sequence can be estimated by

$$\hat{k}^* = \min\left\{k: \|F_{n,k}^0\|_\mathcal{B} = \max_{1 \leq j \leq n}\|F_{n,j}^0\|_\mathcal{B}\right\} \quad (10)$$

Also similar to that in ordinary functional data setting [46], the significance test for the mean break can be realized based on the limiting distribution of the following statistic:

$$T_{f,n} = \max_{1 \leq k \leq n}\|F_{n,k}^0\|_\mathcal{B}^2 \quad (11)$$

Determining the limiting distribution of the test statistic $T_{f,n}$ directly in the Bayes space is not straightforward. Fortunately, the Bayes space $\mathcal{B}^2([0,1])$ has an attractive property that it is isometrically isomorphic to the $L^2([0,1])$ space [54, 55], see the Appendix for more details. One mapping that can serve as the isometric isomorphism between $\mathcal{B}^2([0,1])$ and $L^2([0,1])$ is the centered log-ratio (clr) transformation defined as follows [54, 55]:

$$\text{clr}[f](x) = \log f(x) - \int_0^1 \log f(\tau) d\tau, \quad x \in [0,1], \; f \in \mathcal{B}^2([0,1]) \quad (12)$$

In the sense of isometric isomorphism, the test statistic given in Eq. (11) can be equivalently written as

$$T_{f,n} = \max_{1 \leq k \leq n}\|F_{n,k}^0\|_\mathcal{B}^2 = \max_{1 \leq k \leq n}\|\text{clr}[F_{n,k}^0]\|_2^2 = T_{\text{clr}[f],n} \quad (13)$$

where $\|\cdot\|_2$ stands for the $L^2$-norm defined as $\|u\|_2 = \left(\int_0^1 u^2(\tau) d\tau\right)^{1/2}$, $\forall u \in L^2([0,1])$, $\text{clr}[F_{n,k}^0]$ is the clr-transformation of the FCSS $F_{n,k}^0$ given in Eq. (9). Using the property given in Eq. (25) of the Appendix, the expression of $\text{clr}[F_{n,k}^0]$ can be written as

$$\text{clr}[F_{n,k}^0](x) = \frac{1}{\sqrt{n}}\left(\sum_{i=1}^{k}\text{clr}[f_i](x) - \frac{k}{n}\sum_{i=1}^{n}\text{clr}[f_i](x)\right) \tag{14}$$

It is worth noting that the clr-transformed result $\text{clr}[F_{n,k}^0](x)$ also has a constraint of integrating to zero [55], thereby $\text{clr}[F_{n,k}^0]$ actually resides in a subspace of $L^2([0,1])$. If we neglect this constraint, namely the clr-transformed functions are treated as ordinary functions in $L^2([0,1])$), then the limiting distribution of the test statistic given in Eq.(13) under $H_0$ is [46]

$$T_{\text{clr}[f],n} = \max_{1\leq k\leq n}\|\text{clr}[F_{n,k}^0]\|_2^2 \xrightarrow{d} \sup_{0\leq x\leq 1}\sum_{l=1}^{\infty}\lambda_l B_l^2(x) = T_s, \quad n\to\infty, x\in[0,1] \tag{15}$$

where $\xrightarrow{d}$ stands for convergence in distribution, $B_l(x)$s are standard Brownian bridges defined on $[0,1]$ and independent across $l$, $\lambda_l$s are the eigenvalues of the covariance operator associated with the clr-transformed error sequence $\{\text{clr}[\varepsilon_i](x): 1\leq i\leq n\}$ (the calculation will be provided later). Since the clr-transformed results belong to a subspace of $L^2([0,1])$, $T_s = \sup_{0\leq x\leq 1}\sum_{l=1}^{\infty}\lambda_l B_l^2(x)$ given in Eq. (15) is an approximation to the true limiting distribution of the test statistic $T_{\text{clr}[f],n}$. Such an approximation may introduce errors to the hypothesis testing in real applications; however, related simulation studies as well as real data applications conducted later in this article indicate that this approximated limiting distribution is effective in detecting the change-points of the investigated distributional sequences. Moreover, related simulation studies will also validate that such a clr-transforming strategy is superior to an alternative strategy that directly regarding PDFs as ordinary functions.

We now turn to discussing how to compute the eigenvalue sequence $\{\lambda_l: l=1,2,\cdots\}$ involved in Eq. (15). According to the related theory of functional data analysis [22, 46], the covariance kernel $C_{\text{clr}[\varepsilon_f]}$ and the covariance operator $c_{\text{clr}[\varepsilon_f]}$ associated with the clr-transformed error sequence $\{\text{clr}[\varepsilon_i](x): 1\leq i\leq n\}$ are defined as

$$C_{\text{clr}[\varepsilon_f]}(t,s) = \frac{1}{n}\sum_{i=1}^{n}\text{clr}[\varepsilon_i](t)\text{clr}[\varepsilon_i](s), \quad (t,s)\in[0,1]\times[0,1] \tag{16a}$$

$$c_{\text{clr}[\varepsilon_f]}(\eta) = \int C_{\text{clr}[\varepsilon_f]}(\cdot,s)\eta(s)\,ds, \quad \eta\in L^2([0,1]), \ s\in[0,1] \tag{16b}$$

The covariance operator $c_{\text{clr}[\varepsilon_f]}$ is an integral operator induced by the covariance kernel $C_{\text{clr}[\varepsilon_f]}$. Performing an eigenvalue decomposition of the covariance operator $c_{\text{clr}[\varepsilon_f]}$ can yield the eigenvalue sequence $\{\lambda_l: l = 1,2,\cdots\}$ in descending order, and the corresponding eigenfunctions satisfy

$$C_{\text{clr}[\varepsilon_f]}(\varphi_l)(x) = \lambda_l \varphi_l(x), \qquad l = 1, 2, \cdots \tag{17}$$

For computational details about such an eigenvalue decomposition of covariance operator associated with functional data, see e.g. the monograph of Ramsay and Silverman [22]. In practical applications, the error sequence is unknown, it has to be calculated from the clr-transformed functions $\{\text{clr}[f_i]\}$ by subtracting the sample mean function. Recall that the clr-transformed functions are treated as ordinary functional data in this study; thus, the calculation for the empirical covariance kernel $\hat{C}_{\text{clr}[\varepsilon_f]}$ using the clr-transformed functions $\{\text{clr}[f_i]: i = 1,2,\cdots,n\}$ is the same as that in ordinary functional data setting, see e.g., Aue et al. [43, 46] and Aston and Kirch [45]. For computational convenience, the infinite eigenvalue sequence $\{\lambda_l: l = 1,2,\cdots\}$ involved in the limiting distribution $T_s$ is recommended to be truncated at $l = L$ with $L$ determined by $L = \min\left\{l^\# \in \mathbb{N}: \frac{\sum_{l=1}^{l^\#} \lambda_l}{\sum_{l=1}^{\infty} \lambda_l} \geq \theta\right\}$, where $\theta$ is a given threshold and its default value is set to 0.95 throughout this study.

With the approximated limiting distribution of the test statistic at hand, the hypothesis testing for judging whether $\delta_f = 0$ or not can be realized by checking whether the calculated test statistic $T_{f,n}$ falls into the rejection region of the limiting distribution $T_s$ at a pre-specified significance level $\alpha \in (0, 1)$. However, directly determining the rejection region is not straightforward, a Monte Carlo-based approach is more practical [46]. Specifically, the i.i.d. Brownian bridges $\{B_l(x): l = 1, 2, \cdots, L\}$ involved in the truncated limiting distribution $T_{s,L} = \sup_{0 \leq x \leq 1} \sum_{l=1}^{L} \lambda_l B_l^2(x)$ can be first simulated by the Monte Carlo method implemented in the R package *sde* (*https://cran.r-project.org/*); then, by plugging the simulated $\{B_l(x): l = 1, 2, \cdots, L\}$ into $T_{s,L}$, one can obtain a Monte Carlo sample of $T_{s,L}$. Repeat the above procedure for $M$ times and denote the resulting Monte Carlo samples as $\{T_{s,L}^1, T_{s,L}^2, \cdots, T_{s,L}^M\}$; then, the p-value associated with the hypothesis testing can be calculated by

$$p = \frac{1}{M} \sum_{k=1}^{M} \chi_{[T_{f,n}, \infty)}(T_{s,L}^k) \tag{18}$$

where $T_{f,n}$ is the test statistic calculated by Eq.(13), and $\chi_A(\cdot)$ is the indicator function. If the p-value is less than the pre-specified significance level, then it suggests that the test statistic $T_{f,n}$ falls into the rejection region of the limiting distribution; thus, $H_0$ should be rejected.

*2.4. Measures for robustness improvement under data contamination*

The FCSS defined in Eq. (9) plays a critical role in the change-point detection. Such a statistic is constructed by combining the sample mean function of the PDFs coming before $k$ with the global sample mean function of the whole sequence. Since the sample mean is sensitive to outliers, the calculated FCSS may also be seriously distorted by the outlying PDFs presented in the distributional sequence. For a more reliable detection, effective measures are required to appropriately handle the outlying PDFs. For this purpose, we recommend to conduct a data cleaning to the raw distributional data. Specifically, the potential outlying PDFs hidden in the raw data are first detected by choosing appropriate functional outlier detection methods, and then removed from the distributional sequence before implementing the change-point detection.

Although functional outlier detection methods for ordinary functional data have been well developed, relative research contributions for distributional outlier detection are still quite rare. Most recently, Lei et al. [57] have conducted a systematic study on distributional outlier detection, and relevant computationally efficient detection tools have been developed, which shows good performance in outlying PDF detection. Following the recommendation in Lei et al. [57], the Tree-Distance distributional outlier detection method [57] is selected as the primary tool for outlying PDF detection in this study, and the QF-FDO method [57] is utilized as a complementary tool depending on the specific situations.

Given the distributional sequence denoted as $\mathcal{F} = \{f_1, f_2, \cdots, f_n\}$, the change-point detection procedure after fusing the distributional outlier cleaning processing is outlined as follows:

Step 1: perform the Tree-Distance outlier detection method to $\mathcal{F}$;

Step 2 (optional): perform the QF-FDO outlier detection method to $\mathcal{F}$;

Step 3: remove the outlying PDFs detected above from $\mathcal{F}$, and denote the remaining sub-sequence as $\mathcal{F}^- = \{f_{k_1}, f_{k_2}, \cdots, f_{k_n}\}$, where $k_j$ is the original index of the PDF in the raw sequence;

Step 4: perform change-point detection to $\mathcal{F}^-$ using the method described in Subsection 2.3;

Step 5: convert the change-point location detected in Step 4 to its corresponding location in the original sequence.

For Step 5, support the change-point detected at Step 4 occurs at $f_{k_*}$, then the corresponding change-point in the original sequence is located at $k = k_*$ because the indices of the elements in the sub-sequence $\mathcal{F}^-$ are the same with their original indices used in $\mathcal{F}$.

## 3. Simulation studies

### 3.1 Simulation study I

In this subsection, a simulation study is conducted to validate the effectiveness of the proposed Bayes space-based change-point detection method, as well as to demonstrate its superiority by comparing with an alternative detection strategy.

The distributional sequence is simulated by a two-stage data-generating process: firstly, the following functional change model is used to generate a raw functional sequence denoted as $\{\tilde{f}_i(x)\}_{i=1}^n$; then, $\tilde{f}_i(x)$s are converted into PDFs to yield the desired distributional sequence.

$$\tilde{f}_i(x) = f_{Beta}(x; a_i, b_i) + 0.8 \cdot \chi_{[k^*+1,n]}(i), \qquad x \in [0,1], i = 1, \cdots, n \tag{19}$$

where $f_{Beta}(x; a_i, b_i)$ is the density of a Beta distribution with parameters $a_i$ and $b_i$ generated by

$$a_i \sim U(14, 25), \; b_i = \text{sort}(\{a_i\}_{i=1}^n), i = 1, \cdots, n \tag{20}$$

where $U(u, v)$ stands for the uniform distribution on $[u, v]$, and $\text{sort}(A)$ represents sorting the elements in set $A$ in ascending order. To convert the generated functional sequence to distributional sequence, the data are further processed as follows:

$$f_i(x) = \frac{\tilde{f}_i(x) - h}{\int_0^1 (\tilde{f}_i(\tau) - h) d\tau}, \; h = \min_{1 \le i \le n} \inf_{x \in [0,1]} \{\tilde{f}_i(x) | i = 1, 2, \cdots, n, x \in [0,1]\} \tag{21}$$

The resulting distributional sequence is denoted as $\mathcal{F} = \{f_1, f_2, \cdots, f_n\}$. A representative sample dataset generated by this model ($n = 100$ and $k^* = 50$) is visualized in Fig. 1.

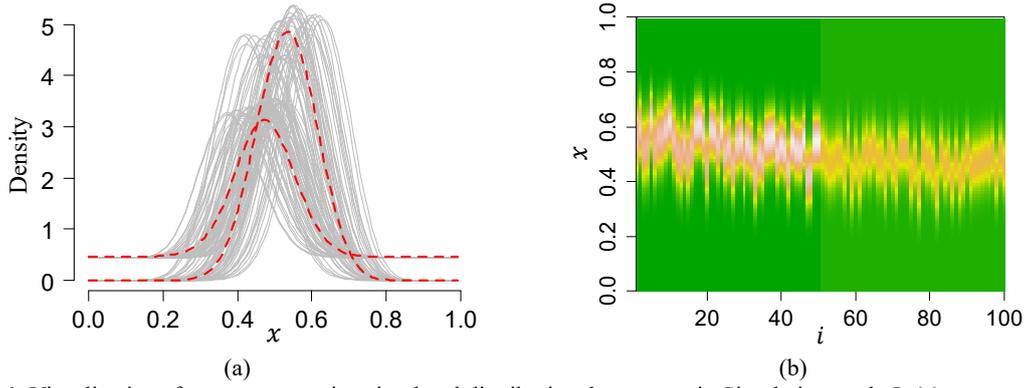

(a) (b)

**Fig. 1.** Visualizations for a representative simulated distributional sequence in Simulation study I: (a) curves of the PDFs and (b) the heatmap representation of the PDF sequence. The two dashed lines in (a) represent the sample mean functions (calculated by using Eq. (6)) of the PDFs coming before and after the change-point.

Two detection methods are considered for performance comparison: (1) the first one is the proposed method; (2) the second one is the competing method, which treats PDFs as ordinary functional data with the change-point detected by using the ordinary functional mean break detection method described in Aue et al.[46]. In the hypothesis testing, the significance level for each method is set to $\alpha = 0.05$, and the parameter $M$ in Eq. (18), corresponding to the number of the Monte Carlo samples of $T_{s,L}$, is set to 2000. The performances of the two methods are evaluated based on their detection accuracies in 500 repeated detection experiments. In each experiment, the distributional sequence is re-generated using the aforementioned data-generating procedure with $n = 100$ and $k^* = 50$. The accuracy of the change-point estimation is quantified by the absolute error defined as $e = |\hat{k}^* - k^*|$, where $\hat{k}^*$ represent the estimate of $k^*$. To avoid the numerical issue encountered in the logarithmic computation involved in the clr-transformation (see Eq. (12)) if the PDF takes values of zero, all the PDFs are processed by $f_i(x) = 0.9 f_i(x) + 0.1$, $x \in [0,1]$. Unless otherwise stated, this processing will be conducted by default throughout the rest of this study. In all the 500 repeated experiments, the calculated p-values associated with the two methods are all less than the pre-specified significance level (i.e., $\alpha = 0.05$). The absolute errors of the change-point locations estimated by the two methods in the 500 repeated experiments are displayed in Fig. 2 (a) and summarized as the boxplots shown in Fig. 2 (b). Comparing the results, one can see that the proposed method significantly outperforms the competing method, indicating that the mean break model constructed in the Bayes space is more suitable for PDF-valued data.

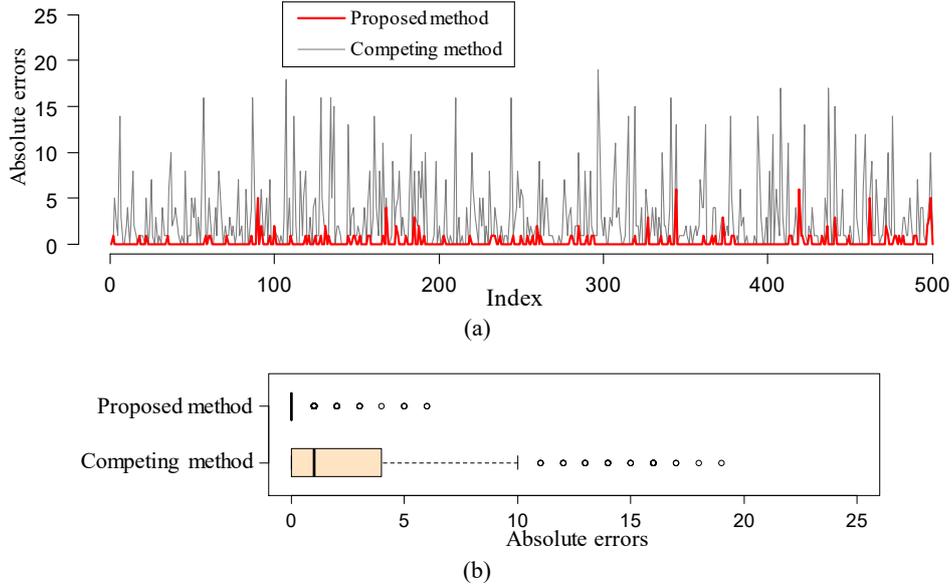
**Fig. 2.** Visualizations for the absolute errors of the change-point location estimates (in the 500 repeated experiments) associated with the two methods in Simulation study I: (a) the error sequences and (b) the boxplots (of the errors).

*3.2 Simulation study II*

This simulation study aims at investigating the impacts of outlying PDFs on the distributional change-point detection, as well as validating the effectiveness of the distributional outlier cleaning strategy in ensuring the detection quality under data contamination.

The following three different data-generating models are considered to produce the synthetic data:

Model I: before the change-point (i.e., $1 \leq i \leq k^*$), the distributional data are generated by a Beta distribution model $Beta(a_i, b_i)$ with parameters $a_i$ and $b_i$ independently generated by $a_i \sim U(10,15), b_i \sim U(10,15)$; after the change-point (i.e., $k^* + 1 \leq i \leq n$), the distributional data are generated by a mixture Beta distribution model $0.5Beta(\tilde{a}_i, \tilde{b}_i) + 0.5Beta(a_i^\#, b_i^\#)$ with the parameters $\tilde{a}_i$, $\tilde{b}_i$, $a_i^\#$ and $b_i^\#$ independently generated by $\tilde{a}_i \sim U(25,40)$, $\tilde{b}_i \sim U(15,20)$, $a_i^\# \sim U(2,4)$ and $b_i^\# \sim U(4,6)$. The resulting distributional sequence is denoted as $\mathcal{F}_I = \{f_i^I(x): i = 1, \cdots, n\}$. Fig. 3 visualizes the representative sample data generated by this model with $n = 100$ and $k^* = 50$.

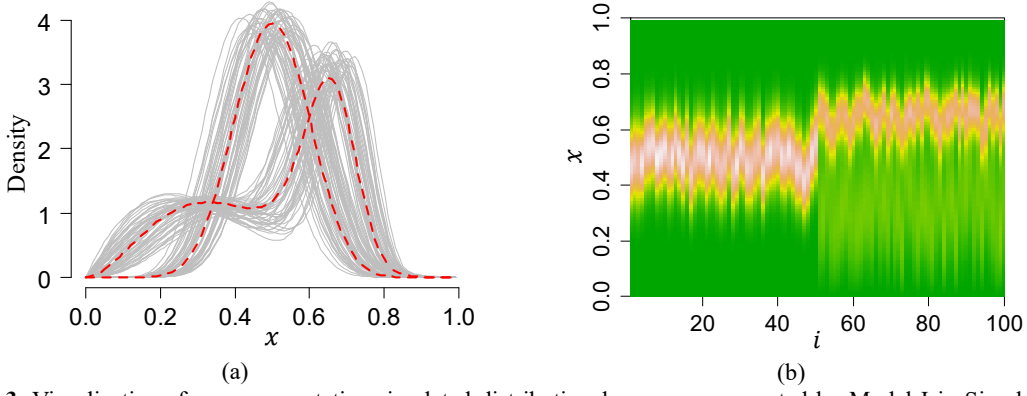

**Fig. 3.** Visualizations for a representative simulated distributional sequence generated by Model I in Simulation study II: (a) curves of the PDFs and (b) the heatmap representation of the PDF sequence. The two dashed lines in (a) represent the sample mean functions (calculated by using Eq. (6)) of the PDFs coming before and after the change-point.

Model II: the distributional sequence is generated by a Beta distribution model with the parameters following different distributions before and after the change-point. Specifically, before the change-point, the parameters $a_i$ and $b_i$ of the Beta distribution $Beta(a_i, b_i)$ are generated by $a_i \sim U(15,25)$ and $b_i = \beta a_i$ with $\beta = \frac{1}{c} - 1$; after the change-point, the parameters of the Beta distribution are generated by $\tilde{a}_i \sim U(5,10)$ and $\tilde{b}_i = \beta \tilde{a}_i$ with $\beta = \frac{1}{c} - 1$. The resulting distributional sequence can be written in a unified form as follows:

$$f_i^{II}(x) = \chi_{[1,k^*]}(i) f_{Beta}(x; a_i, b_i) + \chi_{[k^*+1,n]}(i) f_{Beta}(x; \tilde{a}_i, \tilde{b}_i) \tag{22}$$

where $f_{Beta}(x; a, b)$ stands for the density of the Beta distribution with parameters $a$ and $b$. According to the theory of Beta distribution, the parameter $c$ involved in the above data-generating procedure is actually the expectation (i.e., the first-order moment) of the random variable $X \sim Beta(a, b)$ with $b = \beta a = (\frac{1}{c} - 1)a$, i.e., $c = \int x f_{Beta}(x; a, b) \, dx$. Therefore, the underlying random variables associated with the distributions generated by Model II have the same expectations. Throughout this simulation study, the value of $c$ is fixed at 0.45, the resulting distributional sequence is denoted as $\mathcal{F}_{II} = \{f_i^{II}(x): i = 1, \cdots, n\}$ and the corresponding representative sample data ($n = 100$ and $k^* = 50$) are visualized in Fig. 4.

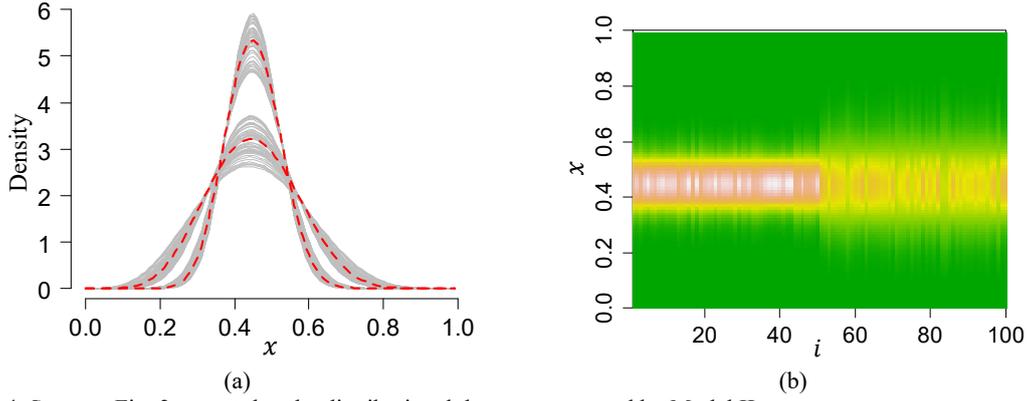
**Fig. 4.** Same as Fig. 3 except that the distributional data are generated by Model II.

Model III: the distributional sequence is also generated by the Beta distribution model, but the parameters are simulated using different principles. Specifically, before the change-point, the parameters $a_i$ and $b_i$ of the Beta distribution are generated by $a_i \sim U(15,25)$ and $b_i = \beta_i a_i$ with $\beta_i$ independently drawn from $U(0.85, 1.0)$; after the change-point, the parameters $\tilde{a}_i$ and $\tilde{b}_i$ of the Beta distribution are generated by $\tilde{a}_i \sim U(15,25)$ and $\tilde{b}_i = \tilde{\beta}_i \tilde{a}_i$ with $\tilde{\beta}_i$ independently drawn from $U(1.0 + q, 1.15 + q)$, where $q \sim U(0.005, 0.015)$. Then, the desired distributional sequence can be obtained by substituting the parameters of the Beta distributions generated above into Eq. (22), the result is denoted as $\mathcal{F}_{III} = \{f_i^{III}(x): i = 1, \cdots, n\}$ and the corresponding representative sample data ($n = 100$ and $k^* = 50$) are visualized in Fig. 5.

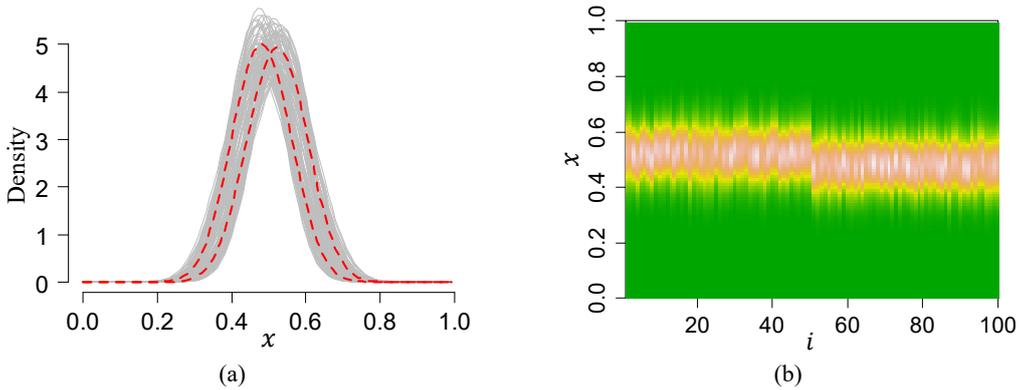
**Fig. 5.** Same as Fig. 3 except that the distributional data are generated by Model III.

Comparing the simulated distributional sequences generated by the three Models shown in Figs.3-5, the PDF-valued data associated with the first two models can be viewed as the strong-change case since the corresponding mean functions (represented by the dashed bold lines) have changed dramatically after the change-point. In contrast, the mean function of the PDFs generated by Model III only exhibits a slight change after the change-point; thus, it can be viewed as the mild-change case. Moreover, one can see from Fig. 4 that the PDFs generated by Model II are aligned functional data, because the underlying Beta distributed random variables share the same expectation. It is worth noting that the expectation, namely the mean of the underlying random variable, is different from the mean function of the PDFs calculated by Eq. (6). Such a model is to simulate the fact that the monitoring data may share the same mean but follow different distributions. One can learn from this model that the changes in the distributions of the monitoring data cannot be detected by merely using the scalar sample mean of the raw measurements (belong to scalar data) in some situations, because the scalar sample mean only reflects the information of the first-order moment of the distribution. In contrast, in our proposed functional data analytic-based distributional change-point detection method, the PDFs themselves are directly analyzed as the random objects; thus, more complete information of the probability distributions can be preserved for investigation. In practical applications, the mild-change case represented by Model III is analogous to the distributional information break of monitoring data caused by a minor condition change of the structural system. In this case, the abrupt change in the distributions is not easy to be discovered by manual inspection, whereas the proposed method can provide a more effective tool for detecting such a minor change hidden in the distributional sequence because the decision is made according to the result of hypothesis testing at a given significance level.

---

**Algorithm 1** Generate $N_o$ outlying PDFs

---

Set $\text{Out}_{\text{PDF}} = \emptyset$
for $i = 1$ to $N_o$ do
    Generate $z \sim U(0,1)$
    **if** $z > 0.7$ **then**
        Generate $\mu_1 \sim U(0.3, 0.4)$, $\mu_2 \sim U(0.6, 0.7)$, $a_1 \sim U(8, 14)$, $a_2 \sim U(15, 20)$
        Compute $f_i^o(x) = 0.5 \cdot f_{Beta}\left(x; a_1, \frac{a_1}{\mu_1} - a_1\right) + 0.5 \cdot f_{Beta}\left(x; a_2, \frac{a_2}{\mu_2} - a_2\right)$
    **else**
        Generate $y \sim U(0,1)$, $a \sim U(2,5)$, $b \sim U(13,16)$, $c \sim U(17,22)$ and $d \sim U(2,5)$
        Compute $f_i^o(x) = f_{Beta}(x; a, b) \cdot \chi_{(0.5,1]}(y) + f_{Beta}(x; c, d) \cdot \chi_{[0,0.5]}(y)$
    **end if**
    Set $\text{Out}_{\text{PDF}} \leftarrow f_i^o(x)$
end for
Output $\text{Out}_{\text{PDF}} = \{f_i^o(x)\}_{i=1}^{N_o}$

To simulate data contamination, the procedure outlined in Algorithm 1 is firstly used to generate $N_o$ outlying PDFs; then, $N_o$ elements in the raw distributional sequence are randomly selected to be replaced by the generated outlying PDFs. For notation simplification, the three contaminated distributional sequences are still denoted as $\mathcal{F}_I$, $\mathcal{F}_{II}$ and $\mathcal{F}_{III}$.

Next, the proposed distributional change-point detection method is applied to the contaminated distributional sequences associated with Models I-III to investigate its performance in resisting the effects of outlying PDFs. For each of the three models, the number of PDFs in the simulated distributional sequence is set to 100, the change-point location $k^*$ is fixed at 50, and the outlier contamination rate is set to $a = 20\%$ (i.e., $N_o = 20$). In the change-point detection, the same argument setting as that in Simulation study I is adopted. The above detection experiments are repeated for 500 times with the distributional data re-generated in each repetition. The absolute errors of the estimated change-point locations (i.e., $e = |\hat{k}^* - k^*|$) in the 500 repeated experiments are summarized as the boxplots displayed in Fig. 6 (a) for the three data-generating models. For comparison, the absolute errors of the estimated change-point locations obtained by using the uncontaminated data (i.e., the raw simulated distributional data before inserting the synthetic outlying PDFs) are also presented as the boxplots shown in Fig. 6 (b). Comparing the results, one can see that the detection results associated with the three models have all been affected by the outlying PDFs, because a portion of the estimated change-points using the contaminated data have deviated from the true change-points especially for the mild-change case (i.e., Model III). In the significance tests, the calculated p-values associated with Models I and II are all less than the pre-specified significance level (i.e., $\alpha = 0.05$) in the 500 repeated experiments, suggesting that all the detected change-points in the two strong-change cases have passed the significance tests (meaning that the distributional sequences have change-points at the given significance level). However, for Model III, a certain number of the detected change-points using the contaminated data fail to pass the corresponding significance tests. Obviously, this result is not consistent with the facts that all the distributional sequences generated by Model III did exist mean breaks, the invalid detections are mainly attributed to the interference of data contamination. From these results, one can conclude that the mild-change case is more sensitive to the outlying PDFs than the two strong-change cases.

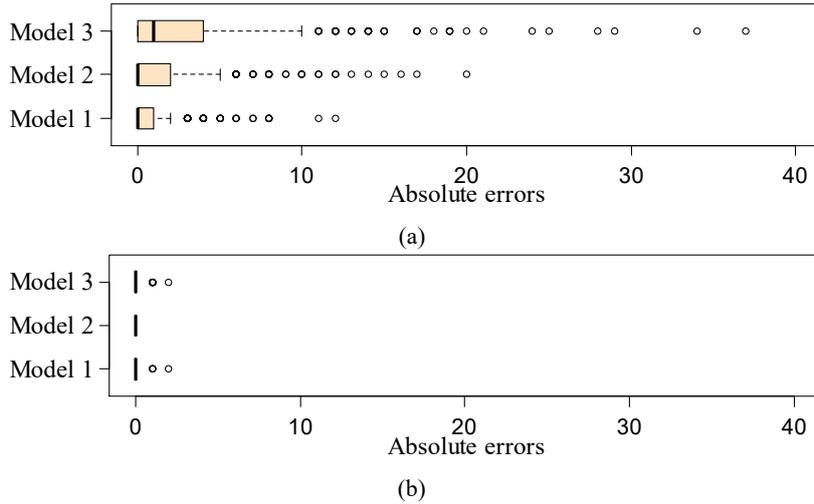

**Fig. 6.** Boxplots of the absolute errors of change-point location estimates associated with Models I-III in the 500 repeated experiments: (a) using the contaminated distributional data and (b) using the uncontaminated distributional data.

The above simulation results indicate that effective measures are generally required to deal with the outlying PDFs; otherwise the detection results might be deceptive, especially for the case of mild-change. To address this issue, we have recommended the distributional outlier cleaning strategy as described in Subsection 2.4. Next, the contaminated distributional data associated with Models I-III simulated above will further be utilized to evaluate the performance of the proposed change-point detection method after fusing the distributional outlier cleaning strategy. The Tree-Distance distributional outlier detection method proposed by Lei et al. [57] is employed to detect and remove the outlying PDFs using the default parameter settings (listed in Table A-2 of the online supplement in Lei et al. [57]). After implementing the distributional outlier cleaning, the accuracies of the change-point detection in the 500 repeated experiments are displayed in Fig. 7. Compared to their counterparts shown in Fig. 6 (a), one can see that the absolute errors of the change-point location estimates have been concentrated around zero, similar to the case without contamination shown in Fig. 6 (b). Moreover, after implementing the distributional outlier cleaning, the calculated p-values in the 500 repeated experiments associated with the three Models are all less than the pre-specified significance level of $\alpha = 0.05$. Such results suggest that the alternative hypothesis (i.e., there exist change-point in the distributional sequence) in each experiment should be accepted at the given significance level. In other words, the outlier-induced invalid detections encountered earlier have disappeared after distributional outlier cleaning.

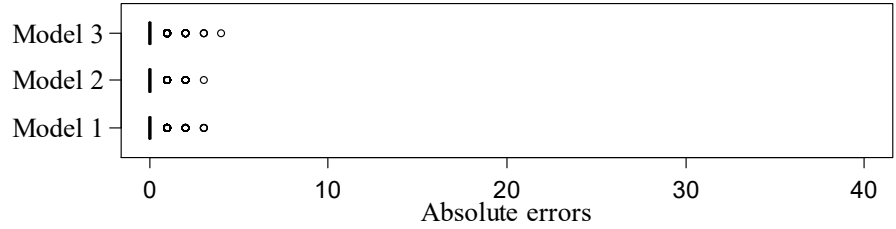

**Fig. 7.** Boxplots of the absolute errors of change-point location estimates associated with Models I-III in the 500 repeated experiments after performing the distributional outlier cleaning.

## 4. Application in structural health monitoring

In this section, the proposed distributional change-point detection method is applied to detect the existence of abrupt change in the distributions of the DSF data which are extracted from the cable tension monitoring data of a long-span cable-stayed bridge. Researchers have found that the ratio of the cable tensions associated with a cable pair located in the same cross-section of the bridge deck is a feature that is sensitive to the change of the cables' health condition but less sensitive to the change of external loads [24]. The abrupt changes in the distributions of the cable-tension ratios can provide critical information in diagnosing the changes in the structural conditions of stay cables. Actually, the information of abrupt change in distributions of extracted DSFs of monitoring data plays an important role in SHM for data-driven structural condition assessments, especially for some pattern recognition-based diagnosis methods [2]. In bridge engineering, except for stay cables, similar distribution change-based structural condition assessment method has also been successfully applied in diagnosing the local health conditions of the bridge's steel deck [25]. To date, effective methods for automatically detecting such distributional changes have not been well developed in SHM. It is worth noting that automatic diagnosis is of crucial interest for practical engineering applications, especially for large scale structural systems. Take the cable-stayed bridge investigated in Li et al. [24] for an example, the cable system is composed of 84 pairs of cables; it is impractical to examine the changes in the distributions of the DSF data by manual inspection for all cable pairs during the whole monitoring period, whereas the distributional change-point detection method developed in this study has good potential for such applications.

In this real data study, the investigated data are also the cable tension monitoring data of the long-span cable-stayed bridge described in Li at al. [24]. The main axis of the bridge is north-south directed, the 84 pairs of cables are symmetrically located at the edges of the upstream and downstream sides of the bridge (see Fig. 1 in [24]). In this study, for cable pair localization, the 84 pairs of cables are labeled as CP-1 to CP-84 from the south to north, namely, the southernmost located cable pair is referred to as CP-1, whereas the northernmost located cable pair is referred to as CP-84. The cable-force monitoring data associated with three representative cable pairs indexed by CP-67, CP-26 and CP-19 are selected for investigation. For each cable pair, 300 days of monitoring data are extracted from the time periods listed in the second row of Table 1 to calculate the cable-tension ratio defined in Li et al. [24]. Using the same data processing procedures (including the data pre-processing) described in Li et al. [24], the DSF data can be extracted from the raw data. The only difference is that the logarithm computation given in Eq. (13) in Li et al. [24] is abandoned, the main reason is that the distributions in our methods are assumed to be finitely supported on a compact interval $[a, b]$ and the logarithm map will take the data to be valued in the interval of $(-\infty, +\infty)$. Therefore, the extracted DSF data in this study is the exponential of the cable-tension ratio defined in Eq. (13) in Li et al. [24]. However, for convenience, the extracted DSF data is also called the cable-tension-ratio (CTR) data throughout the rest of this study.

**Table 1**

The basic information of the extracted distributional data associated with cable pairs CP-67, CP-26 and CP-19.

| Cable pair | CP-67 | CP-26 | CP-19 |
| --- | --- | --- | --- |
| Time period | Mar., 2008 – Apr., 2012 | Apr., 2008 – Mar., 2012 | Nov., 2007 – Jun., 2010 |
| Number of extracted PDFs ($n$) | 300 | 300 | 300 |
| True change-point ($k^*$) | 162 | 145 | 213 |

Before estimating the distributions, a further data processing consisting of scalar outlier filtration and data normalization is conducted to the CTR data. Take the CTR data of CP-67 as an example to illustrate the procedure, the CTR data of the other two cable pairs, namely CP-26 and CP-19, can be processed in a similar way. Firstly, the classic boxplot is employed to detect and filter out the outliers contained in the raw CTR data. It is worth noting that the outlier handled here is scalar outlier, which is different from the functional outlier described in Subsection 2.4. Fig. 8 displays the comparison for a piece of CTR data before and after the scalar outlier filtration. Secondly, the common support of the distribution (i.e., the domain where the corresponding PDF defined on) followed by the CTR data (To make the estimated common support more general, the common support is estimated using the CTR data from a wider time span (about six years) than that listed in Table 1) is estimated using the support estimation method described in Chen et al. [26, 27], then the CTR data are normalized to [0, 1] through support transformation. Finally, the resulting 300-day CTR data are split into daily segments, and the PDF associated with each day is estimated by using the kernel density estimation technique [27]. The last two steps of data processing are similar to that in Chen et al. [26, 27]. By this processing, a total of 300 PDFs can be extracted from the CTR data for each cable pair, and the PDFs are ordered in time to form the distributional sequence for further investigation. For convenience, the extracted distributional sequences associated with cable pairs CP-67, CP-26 and CP-19 are denoted as $\mathcal{F}_{67} = \{f_i^{67}(x): i = 1, \cdots, 300\}$, $\mathcal{F}_{26} = \{f_i^{26}(x): i = 1, \cdots, 300\}$ and $\mathcal{F}_{19} = \{f_i^{19}(x): i = 1, \cdots, 300\}$, respectively. Fig. 9 visualizes the functional curves (first column) and the heatmaps (second column) of the distributional sequences. In the heatmap, the horizontal axis corresponds to the indices of the PDFs contained in the distributional sequence, and the vertical axis corresponds to the independent variable $x$ of the PDF.

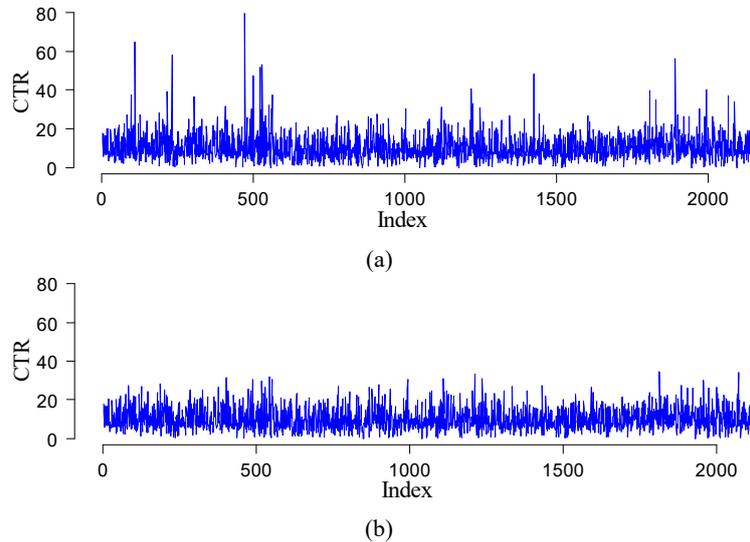

**Fig. 8.** Visualizations for a piece of the raw CTR data associated with cable pair CP-67: (a) before scalar outlier filtration and (b) after scalar outlier filtration.

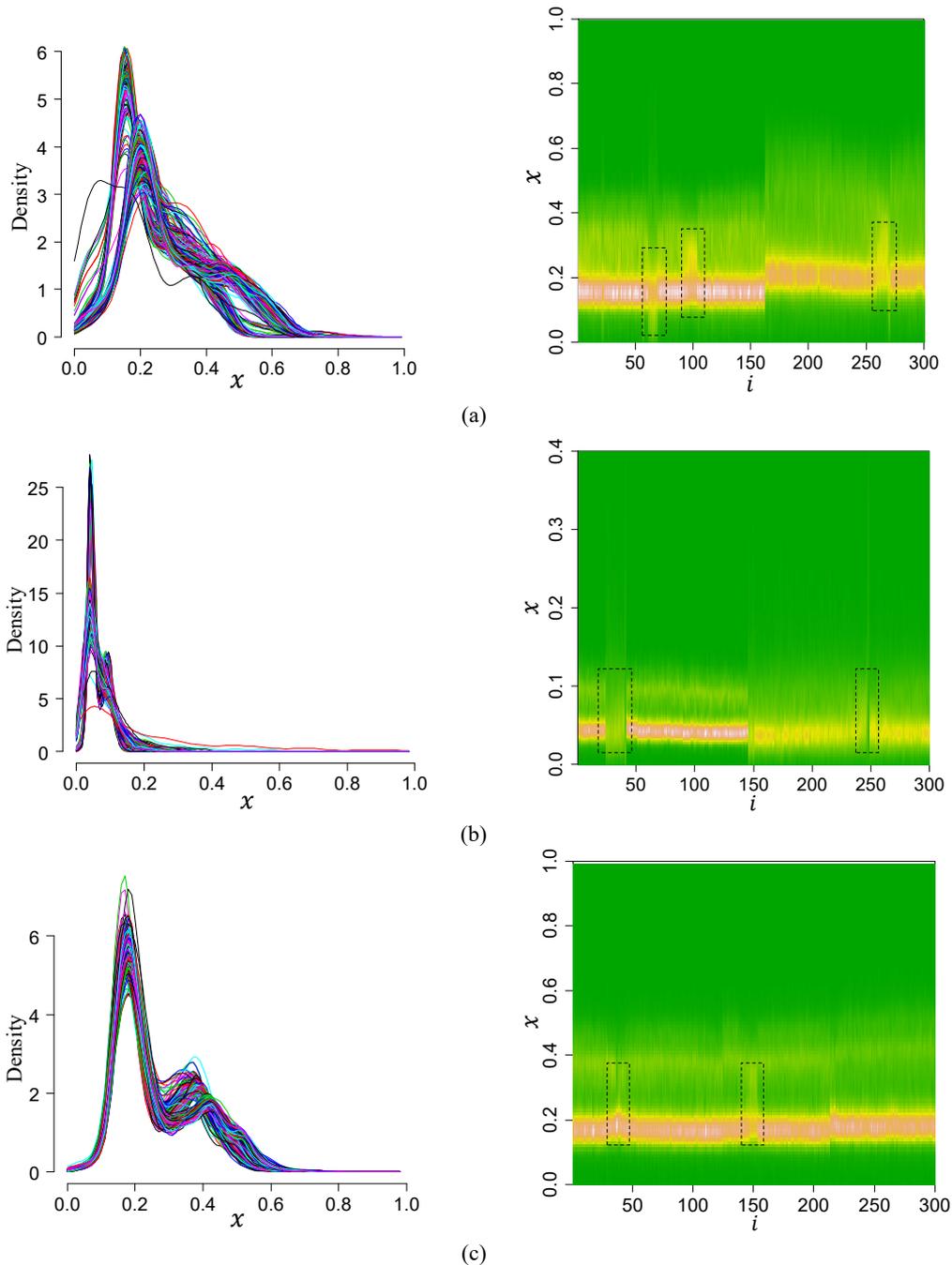

**Fig. 9.** Visualizations for the extracted distributional sequences of the CTR data: (a) the distributional sequence associated with CP-67, (b) the distributional sequence associated with CP-26, and (c) the distributional sequence associated with CP-19. The first column corresponds to the curves of the PDFs, and the second column corresponds to the heatmap representation of the PDF sequence. The dashed rectangle marks the region where the potentially outlying PDFs may present (i.e., the data that possess a feature that distinctly incongruous with the majority of the data).

One can see from the right panel of Fig. 9 (a) that the heatmap of the distributional sequence associated with CP-67 exhibits a significant dislocation near the middle position, indicating that a sharp change has occurred to the PDF-valued sequence. For CP-26, the colors of the heatmap shown in the right panel of Fig. 9 (b) are obviously different between data coming before and after the middle position, which is mainly attributed to the heights of the PDF-curves have reduced significantly after the middle time of the investigated period. Such a phenomenon is similar to that exhibited in the data generated by Model II in Simulation study II. For CP-19, a slight dislocation, located near $i = 200$, can be observed in the corresponding heatmap shown in right panel of Fig. 9 (c). Analogous to Simulation study II, the extracted distributional sequences associated with the first two cable pairs, namely CP-67 and CP-26, can be viewed as the strong-change case, whereas the one associated to CP-19 can be viewed as the mild-change case. Further manual inspection finds that the true change-points in the three distributional sequences are located at $i = 162$, $i = 145$ and $i = 213$ for CP-67, CP-26 and CP-19, respectively, and the results are reported in the fourth row of Table 1. Moreover, all the three distributional datasets contain a certain number of outlying PDFs, see e.g. the data presented in the dashed rectangles of Fig. 9 that possess a feature distinctly incongruous with the majority of the data.

Next, the proposed distributional change-point detection method is employed to detect and locate the change-points contained in the three distributional sequences of CTR data. The following two detection schemes are considered:

Scheme I: do not perform distributional outlier cleaning before change-point detection;

Scheme II: perform distributional outlier cleaning before change-point detection.

Scheme II corresponds to the detection strategy illustrated in Steps 1-5 in Subsection 2.4. Here, the distributional outlier detection methods in Step 1 and Step 2 are both considered. For the Tree-Distance detection method in Step 1, the argument setting is the same as that in Simulation study II; for the QF-FDO outlier detection method (see [57] for the implementation details) in Step 2, the detection region is selected as [0.2, 0.8], and the whisker parameters of the boxplot-based detectors associated with the MO- and VO-directions are set to 1.5 and 2.5, respectively.

In the change-point detection, the argument setting is the same as that in Simulation study I. The detection results are reported in Table 2. In the significance tests, the calculated p-values associated with the three distributional sequences are all less than the pre-specified significance level (i.e., $\alpha = 0.05$) for both the detection schemes, suggesting that the distributional sequences contain change-points (i.e., the null hypothesis $H_0$ should be rejected). Comparing the estimated locations of the change-points, one can see that the results associated with CP-67 and CP-26 are both consistent with the true locations; however, for CP-19, the change-point identified by the first detection scheme (i.e., the one without distributional outlier cleaning) significantly deviates from the true location, whereas the one identified by the second detection scheme is very close to the true location.

**Table 2**

The change-point detection results of the extracted distributional sequences associated with cable pairs CP-67, CP-26 and CP-19.

| Cable pair | CP-67 | CP-26 | CP-19 |
|---|---|---|---|
| True change-point ($k^*$) | 162 | 145 | 213 |
| Estimated change-point ($\hat{k}^*$) | Scheme I: 162<br>Scheme II: 162 | Scheme I: 145<br>Scheme II: 145 | Scheme I: 185<br>Scheme II: 211 |
| Hypothesis testing | Scheme I: Reject $H_0$<br>Scheme II: Reject $H_0$ | Scheme I: Reject $H_0$<br>Scheme II: Reject $H_0$ | Scheme I: Reject $H_0$<br>Scheme II: Reject $H_0$ |

The real data application illustrated above further validates the effectiveness of the proposed distributional change-point detection method as well as demonstrates its practical utility in structural health monitoring. In addition to providing an automatic detection tool for testing the existence of abrupt changes in the distributions of the extracted damage-sensitive features of SHM data, it can also make a more objective judgement on the existence of a change through the hypothesis testing.

## 5. Conclusions

This paper proposed a functional data-analytic method for detecting the change-points occurred in the density-valued sequence extracted from sensor monitoring data, and it is applied to detect and test the distributional information break involved in data-driven structural condition assessment of infrastructure. The effectiveness of the proposed method has been validated by both synthetic and real monitoring data. The following conclusions can be drawn:

(1) The proposed distributional change-point model is constructed using the linear structure of the Bayes space after embedding the PDFs into the Bayes space. Such a model not only has an advantage in the sense of interpretability but also shows better detection accuracy. In the simulation study, the proposed Bayes space-based detection method significantly outperforms a competing method that directly treats PDFs as ordinary functional data.

(2) The proposed distributional change-point detection method can provide an effective tool for automatically diagnosing the information break occurred in the distributions of the DSF data extracted from the measurements of structural responses. The decision about the existence of an abrupt change in the distributional sequence is based on the results of hypothesis testing at a given significance level, thus it can provide a more rational judgement especially when the change is small and indistinguishable by manual inspection.

(3) The change-point detection for the strong-change case is less sensitive to the outlying PDFs presented in the raw distributional data, whereas the detection results of the mild-change case can be seriously affected. The distributional outlier cleaning strategy is a simple and effective measure to cope with the outlying PDFs, relative simulation and real data studies have demonstrated that it can significantly improve the detection performance under data contamination.

(4) In practical applications, it is hard for one to prejudge whether the investigated distributional sequence belongs to the strong-change case or the mild-change case, thus the distributional outlier cleaning is recommended to be performed for all cases.

## Acknowledgments

This work was financially supported by the National Natural Science Foundation of China (Grant Nos. 51908166), the National Key R&D Program of China (Grant No. 2018YFB1600202), China Postdoctoral Science Foundation (Grant No. 2019M661287), and Postdoctoral Science Foundation of Heilong Jiang province.

## Appendix: Introduction to the Bayes space

A detailed introduction to the Bayes space is beyond the scope of this article, this appendix only provides the basic notions and properties related to this study. Readers can refer to Egozcue et al. [53], Van den Boogaart et al. [54] and Hron et al. [55] for more detailed descriptions of the Bayes space.

The Bayes space, denoted as $\mathfrak{B}^2([a, b])$, is a functional space consisting of functions given by

the following set:

$$\mathfrak{B}^2([a,b]) = \left\{f: [a,b] \to \mathbb{R} \middle| f(x) > 0, \forall x \in [a,b] \text{ and } \int_a^b |\log f(s)|^2 ds < +\infty \right\} \quad (23)$$

where $\mathbb{R}$ denotes the real numbers. $\mathfrak{B}^2([a,b])$ is a linear space with its linear operations taking the same form as in Eq.(4) [54, 55]. It can be easily verified that a PDF defined in the compact interval $[a,b]$ belongs to $\mathfrak{B}^2([a,b])$. $\mathfrak{B}^2([a,b])$ can further become a separable Hilbert space if it is equipped with the following inner product [54, 55]:

$$\langle f, g \rangle_\mathfrak{B} = \frac{1}{2(b-a)} \int_a^b \int_a^b \log \frac{f(t)}{f(s)} \log \frac{g(t)}{g(s)} dt ds, \quad f, g \in \mathfrak{B}^2([a,b]) \quad (24)$$

The inner product can naturally induce a norm denoted as $\|\cdot\|_\mathfrak{B} = \sqrt{\langle \cdot, \cdot \rangle_\mathfrak{B}}$. Using the linear operations defined in Eq. (4), a distance can also be induced for the Bayes space, namely $d_\mathfrak{B}(f,g) = \|f \oplus ((-1) \odot g)\|_\mathfrak{B}$.

$\mathfrak{B}^2([a,b])$ is isometrically isomorphic to $L^2([a,b])$ [54, 55]. In other words, there exists a mapping from $\mathfrak{B}^2([a,b])$ to $L^2([a,b])$, denoted as $T: \mathfrak{B}^2([a,b]) \to L^2([a,b])$, that satisfies the following properties:

$$T(f \oplus g) = T(f) + T(g), \quad \forall f, g \in \mathfrak{B}^2([a,b]) \quad (25a)$$

$$T(c \odot f) = c \cdot T(f), \quad \forall c \in \mathbb{R} \text{ and } \forall f \in \mathfrak{B}^2([a,b]) \quad (25b)$$

$$d_\mathfrak{B}(f,g) = d_{L_2}(T(f), T(g)), \quad \forall f, g \in \mathfrak{B}^2([a,b]) \quad (25c)$$

where $d_{L_2}$ is the $L_2$ distance associated with the $L^2([a,b])$ defined as

$$d_{L_2}(\xi, \eta) = \left( \int_a^b (\xi(\tau) - \eta(\tau))^2 d\tau \right)^{1/2}, \quad \xi, \eta \in L^2([a,b]) \quad (26)$$

One mapping that satisfies the properties given in Eq. (25) is the centered log-ratio (clr) transformation [54, 55] given by

$$\text{clr}[f](x) = \log f(x) - \frac{1}{b-a} \int_a^b \log f(\tau) d\tau, \quad x \in [a,b] \text{ and } f \in \mathfrak{B}^2([a,b]) \quad (27)$$

## References

[1] L. Sun, Z. Shang, Y. Xia, S. Bhowmick, S. Nagarajaiah, Review of bridge structural health monitoring aided by big data and artificial intelligence: from condition assessment to damage detection. J. Struct. Eng. 146(5) (2020) 4020073.

[2] Y. Bao, Z. Chen, S. Wei, Y. Xu, Z. Tang, H. Li, The state of the art of data science and engineering in structural health monitoring. Engineering 5(2) (2019) 234–242.

[3] Q. Han, Q. Ma, J. Xu, M. Liu, Structural health monitoring research under varying temperature condition: a review. J. Civ. Struct. Health. 11(1) (2021) 149–173.

[4] K. Worden, E.J. Cross, On switching response surface models, with applications to the structural health monitoring of bridges. Mech. Syst. Signal Process. 98 (2018) 139–156.

[5] M. Vagnoli, R. Remenyte-Prescott, An ensemble-based change-point detection method for identifying unexpected behaviour of railway tunnel infrastructures. Tunn. Undergr. Sp. Tech. 81 (2018) 68–82.

[6] Y. Liao, A.S. Kiremidjian, R. Rajagopal, C.-H. Loh, Structural damage detection and localization with unknown postdamage feature distribution using sequential change-point detection method. J. Aerospace Eng. 32(2) (2019) 4018149.

[7] S. Mariani, P. Cawley, Change detection using the generalized likelihood ratio method to improve the sensitivity of guided wave structural health monitoring systems. Struct. Health Monit. (2020), https://doi.org/10.1177/1475921720981831.

[8] Y. Li, T. Bao, X. Shu, Z. Gao, J. Gong, K. Zhang, Data-driven crack behavior anomaly identification method for concrete dams in long-term service using offline and online change point detection. J. Civ. Struct. Health 11 (2021) 1449–1460.

[9] A. Ostermann, G. Spielberger, A. Tributsch, Detecting structural changes with ARMA processes. Math. Comp. Model Dyn. 22(6) (2016) 524–538.

[10] F. Pozo, I. Arruga, L.E. Mujica, M. Ruiz, E. Podivilova, Detection of structural changes through principal component analysis and multivariate statistical inference. Struct. Health Monit. 15(2) (2016) 127–142.

[11] J.N. Yang, S. Lin, Identification of parametric variations of structures based on least squares estimation and adaptive tracking technique. J. Eng. Mech. 131(3) (2005) 290–298.

[12] F.N. Catbas, H.B. Gokce, M. Gul, Nonparametric analysis of structural health monitoring data for identification and localization of changes: Concept, lab, and real-life studies. Struct. Health Monit. 11(5) (2012) 613–626.

[13] M. Döhler, L. Mevel, F. Hille, Subspace-based damage detection under changes in the ambient excitation statistics. Mech. Syst. Signal Process. 45(1) (2014) 207–224.

[14] J.P. Santos, A. Orcesi, C. Cremona, P. Silveira, Baseline-free real-time assessment of structural changes. Struct. Infrastruct. E. 11(2) (2015) 145–161.

[15] G.M. Lloyd, M.L. Wang, T.L. Paez, Minimisation of decision errors in a probabilistic neural network for change point detection in mechanical systems. Mech. Syst. Signal Process.13(6) (1999) 943–954.

[16] B.E. Parker Jr, H.A. Ware, D.P. Wipf, W.R. Tompkins, B.R. Clark, E.C. Larson, H.V. Poor, Fault diagnostics using statistical change detection in the bispectral domain. Mech. Syst. Signal Process. 14(4) (2000) 561–570.

[17] G. Lu, Y. Zhou, C. Lu, X. Li, A novel framework of change-point detection for machine monitoring. Mech. Syst. Signal Process. 83 (2017) 533–548.

[18] T. Wang, G.-L. Lu, J. Liu, P. Yan, Adaptive change detection for long-term machinery monitoring using incremental sliding-window. Chin. J. Mech. Eng. 30(6) (2017) 1338–1346.

[19] G. Lu, J. Liu, P. Yan, Graph-based structural change detection for rotating machinery monitoring. Mech. Syst. Signal Process. 99 (2018) 73–82.

[20] D.J. Miller, N.F. Ghalyan, S. Mondal, A. Ray, HMM conditional-likelihood based change detection with strict delay tolerance. Mech. Syst. Signal Process. 147 (2021) 107109.

[21] A. Jablonski, M. Bielecka, A. Bielecki, Unsupervised detection of rotary machine imbalance based on statistical signal properties. Mech. Syst. Signal Process. 167 (2022) 108497.

[22] J.O. Ramsay, B.W. Silverman, Functional Data Analysis, second ed. Springer, New York, 2005.

[23] F. Ferraty, P. Vieu, Nonparametric Functional Data Analysis: Theory and Practice. Springer, New York, 2006.

[24] S. Li, S. Wei, Y. Bao, H. Li, Condition assessment of cables by pattern recognition of vehicle-induced cable tension ratio. Eng. Struct. 155 (2018) 1–15.

[25] S. Wei, Z. Zhang, S. Li, H. Li, Strain features and condition assessment of orthotropic steel deck cable-

supported bridges subjected to vehicle loads by using dense FBG strain sensors, Smart Mater. Struct. 26 (10) (2017) 104007.

[26] Z. Chen, Y. Bao, H. Li, B.F. Spencer Jr., A novel distribution regression approach for data loss compensation in structural health monitoring, Struct. Health Monit. 17(6) (2018)1473–1490.

[27] Z. Chen, Y. Bao, H. Li, B.F. Spencer Jr., LQD-RKHS-based distribution-to-distribution regression methodology for restoring the probability distributions of missing SHM data, Mech. Syst. Signal Process. 121 (2019) 655–674.

[28] Q. Lin, C. Li, Kriging based sequence interpolation and probability distribution correction for gaussian wind field data reconstruction. J. Wind Eng. Ind. Aerod. 205 (2020) 104340.

[29] Q. Lin, C. Li, Nonstationary wind speed data reconstruction based on secondary correction of statistical characteristics. Struct. Control Health Monit. (2021) 28(9).

[30] Z. Chen, X. Lei, Y. Bao, F. Deng, Y. Zhang, H. Li, Uncertainty quantification for the distribution-to-warping function regression method used in distribution reconstruction of missing structural health monitoring data. Struct. Health Monit. (2021), https://doi.org/10.1177/1475921721993381.

[31] Z. Chen, Y. Bao, Z. Tang, J. Chen, H. Li, Clarifying and quantifying the geometric correlation for probability distributions of inter-sensor monitoring data: A functional data analytic methodology. Mech. Syst. Signal Process. 138 (2020) 106540.

[32] A. Aue, L. Horváth, Structural breaks in time series. J. Time Ser. Anal. 34(1) (2013) 1–16.

[33] M. Csörgö, L. Horváth, Limit Theorems in Change-Point Analysis. John Wiley & Sons, Chichester, 1997.

[34] J. Chen, A.K. Gupta, Parametric Statistical Change Point Analysis: With Applications to Genetics, Medicine, and Finance, second ed. Birkhäuser, Boston, 2012.

[35] C. Beaulieu, J. Chen, J.L. Sarmiento, Change-point analysis as a tool to detect abrupt climate variations. Phil. Trans. R. Soc. A 370 (2012) 1228–1249.

[36] L.F. Robinson, T.D. Wager, M.A. Lindquist, Change point estimation in multi-subject fMRI studies. NeuroImage 49(2) (2010) 1581–1592.

[37] L. Xiong, C. Jiang, C.-Y. Xu, K. Yu, S. Guo, A framework of change-point detection for multivariate hydrological series. Water Resour. Res. 51(10) (2015) 8198–8217.

[38] M. Basseville, A. Benveniste, G. Moustakides, Detection and diagnosis of abrupt changes in modal characteristics of nonstationary digital signals. IEEE Trans. Inform. Theory 32 (1986) 412–417.

[39] M. Basseville, A. Benveniste, G. Moustakides, A. Rougee, Detection and diagnosis of changes in the eigenstructure of nonstationary multi variable systems. Automatica 23 (1987) 479–489.

[40] M. Basseville, I.V. Nikiforov, Detection of Abrupt Changes: Theory and Applications. Prentice Hall, Englewood Cliffs, 1993.

[41] T.L. Lai, Sequential changepoint detection in quality control and dynamical systems. J. Roy. Statist. Soc. Ser. B 57(4) (1995) 613–644.

[42] I. Berkes, R. Gabrys, L. Horváth, P. Kokoszka, Detecting changes in the mean of functional observations. J. Roy. Statist. Soc. Ser. B 71(5) (2009) 927–946.

[43] A. Aue, R. Gabrys, L. Horváth, P. Kokoszka, Estimation of a change-point in the mean function of functional data. J. Multivariate Anal. 100(10) (2009) 2254–2269.

[44] S. Hörmann, P. Kokoszka, Weakly dependent functional data. Ann. Statist. 38(3) (2010) 1845–1884.

[45] J.A.D. Aston, C. Kirch, Detecting and estimating changes in dependent functional data. J. Multivariate Anal. 109 (2012) 204–220.

[46] A. Aue, G. Rice, O. Sönmez, Detecting and dating structural breaks in functional data without dimension reduction. J. Roy. Statist. Soc. Ser. B 80(3) (2018) 509–529.

[47] P. Kokoszka, H. Miao, A. Petersen, H.L. Shang, Forecasting of density functions with an application to cross-sectional and intraday returns. Int. J. Forecasting 35(4) (2019) 1304–1317.

[48] Y. Chen, Z. Lin, H.-G. Müller, Wasserstein regression. J. Amer. Statist. Assoc. (2021), https://doi.org/10.1080/01621459.2021.1956937.

[49] A. Petersen, C. Zhang, P. Kokoszka, Modeling probability density functions as data objects. Economet. Statist.

(2021), https://doi.org/10.1016/j.ecosta.2021.04.004.

[50] O.H.M. Padilla, A. Athey, A. Reinhart, J.G. Scott. Sequential nonparametric tests for a change in distribution: an application to detecting radiological anomalies. J. Amer. Statist. Assoc. 114(526) (2019) 514–528.

[51] L. Horváth, P. Kokoszka, S. Wang, Monitoring for a change point in a sequence of distributions. Ann. Statist. 49(4) (2021) 2271–2291.

[52] P. Dubey, H.-G. Müller, Fréchet change-point detection. Ann. Statist. 48(6) (2020) 3312–3335.

[53] J. Egozcue, J. Díaz-Barrero, V. Pawlowsky-Glahn, Hilbert space of probability density functions based on Aitchison geometry. Acta Math. Sin. (Engl.Ser.) 22 (2006) 1175–1182.

[54] K. Van den Boogaart, J. Egozcue, V. Pawlowsky-Glahn, Bayes Hilbert spaces. Aust. New Zealand J. Stat. 54 (2014) 171–194.

[55] K. Hron, A. Menafoglio, M. Templ, K. Hrůzová, P. Filzmoser, Simplicial principal component analysis for density functions in Bayes spaces, Comput. Statist. Data Anal. 94 (2016) 330-350.

[56] S. Dasgupta, D. Pati, A. Srivastava, A geometric framework for density modelling. https://arxiv.org/abs/1701.05656v1, 2017.

[57] X. Lei, Z. Chen, H. Li, Functional outlier detection for density-valued data with application to robustify distribution to distribution regression. https://arxiv.org/abs/2110.00707, 2021.